\documentclass[prb,showpacs,floatfix,epsfig,twocolumn,superscriptaddress]{revtex4}
\usepackage{graphicx}
\usepackage{dcolumn}

\begin{document}

\newcommand*{\cm}{cm$^{-1}$\,}

%
\title{Doping evolution of chemical potential, spin-correlation gap,
and charge dynamics in Nd$_{2-x}$Ce$_x$CuO$_4$}
%
%

\author{N. L. Wang}
\author{G. Li}
\author{D. Wu}
\affiliation{Beijing National Laboratory for Condensed Matter
Physics, Institute of Physics, Chinese Academy of Sciences,
Beijing 100080, P.~R.~China}

\author{X. H. Chen}
\author{C. H. Wang}
\affiliation{Hefei National Laboratory for Physics Science at
Microscale and Deaprtment of Physics, University of Science and
Technology of China, Hefei 230026, P. R. China}
\author{H. Ding}
\affiliation{Department of Physics, Boston College, Chestnut Hill,
MA 02467, USA}

%
%
%
\begin{abstract}
We report optical reflectivity study on Nd$_{2-x}$Ce$_x$CuO$_4$
over a broad doping range 0$\leq x\leq$0.20. The study reveals a
systematic shift of chemical potential with doping. A pronounced
peak structure specific to the electron-doped cuprate is observed
in optical scattering rate based on the extended Drude model
analysis. The energy scale of the peak could be very different
from that of the peak in conductivity spectrum. We clarify that
the "hot spots" gap probed in photoemission experiments correlates
directly with the sharp suppression feature in scattering rate
rather than conductivity spectra. The gap is associated with the
short-range antiferromagnetic correlation, and disappears in a
manner of fading away with increasing doping and temperature. In
the heavily overdoped region, a dominant $\omega^2$-dependence of
the scattering rate is identified up to very high energy.
\end{abstract}


\maketitle
\section{Introduction}

A central issue in understanding the mechanism of high-temperature
superconductivity is how an antiferromagnetic (AF) insulator
evolves into a superconductor with electron or hole doping.
Although both the electron- and hole-doped high-$T_c$
superconductors share a lot of similarities, including the d-wave
pairing symmetry, the phase diagrams exhibit rather asymmetric
behaviors \cite{Damascelli}. For the hole-doped cuprates, the AF
order disappears with a small amount of carrier concentration, and
the superconducting phase is well separated from the AF ordered
phase; while in the electron-doped cuprates, the AF order is much
more robust with respect to doing, and the AF and superconducting
phases are adjacent to each other or even coexist. The doping
range where the superconducting transition occurs in
electron-doped cuprates is much narrower and the maximum $T_c$ is
much lower than for hole doped cuprates. Angle-resolved
photoelectron spectroscopy (ARPES) experiments indicated that the
hole carriers doped into the parent compound first enter
($\pi/2,\pi/2$) points in the Brillouin zone and produce a Fermi
arc \cite{Yoshida}. By contrast, for the electron-doped case, a
small Fermi surface pocket appears first around ($\pi,0$), and
another one shows up around ($\pi/2,\pi/2$) upon increasing
doping. The two pockets are separated by a gaped region locating
at the intersecting points (so-called "hot spots") of the Fermi
surface and the AF zone boundary \cite{Armitage}.

Optical spectroscopy can probe not only the low-lying intraband
response, but also the interband transitions from occupied to
unoccupied states. It provides supplementary information about the
electronic states as yielded by ARPES, which detects only occupied
states. Previous optical studies on Nd$_{2-x}$Ce$_x$CuO$_4$ (NCCO)
system \cite{Cooper,Arima,Singley,Onose} have uncovered a transfer
of spectral weight from charge-transfer excitations to low
frequencies with doping, and an occurrence of a large pseudogap in
optical conductivity, which is believed to be a reflection of the
gaped region in the Fermi surface seen in ARPES experiment. A
recent doping-dependent optical study further suggested that this
partial gap is associated with an ordered phase which ends on a
quantum critical point at approximately optimal doping
\cite{Zimmers}. In this paper, we present a more systematic
optical study on NCCO single crystals. Our samples cover very
broad doping range from parent compound to overdoped
non-superconducting compound. The study reveals a number of novel
properties, including a systematic shift of chemical potential, a
spin-correlation gap which fades away with doping/temperature with
no clear phase boundary line, and a dominant $\omega^2$-dependence
of the scattering rate up to very high energy in the
non-superconducting overdoped region. We clarify that the ``hot
spots" gap probed in ARPES experiments correlates directly with
the gap in the scattering rate spectrum rather than with the
suppression feature in optical conductivity spectrum.

\section{Experiments}

The NCCO single crystals were grown from a copper-oxide-rich melt
in Al$_2$O$_3$ crucibles over a wide range of Ce concentration
0$\leq x\leq$0.20. The actual Ce concentration was determined by
inductively coupled plasma spectrometry analysis experiments, and
by the energy-dispersive x-ray analysis using a scanning electron
microscopy, respectively. All samples were annealed in flowing
helium for over 10 hours at 900$^o$C to remove the interstitial
oxygen.

The near-normal incident reflectance spectra ($R(\omega$)) were
measured by a Bruker 66v/s spectrometer in the frequency range
from 50 \cm to 25,000 \cm. The crystal with a very shiny surface
was mounted on an optically black cone in a cold-finger flow
cryostat. An \textit{in situ} gold and aluminum overcoating
technique was employed for reflectance measurements\cite{Homes}.
The optical conductivity spectra were obtained from a
Kramers-Kronig transformation of $R(\omega$). We use Hagen-Rubens
relation for the low frequency extrapolation, and a constant
extrapolation to 100,000 \cm followed by a well-known function of
$\omega^{-4}$ in the higher-energy side.

\begin{figure}[t]
\centerline{\includegraphics[width=2.9in]{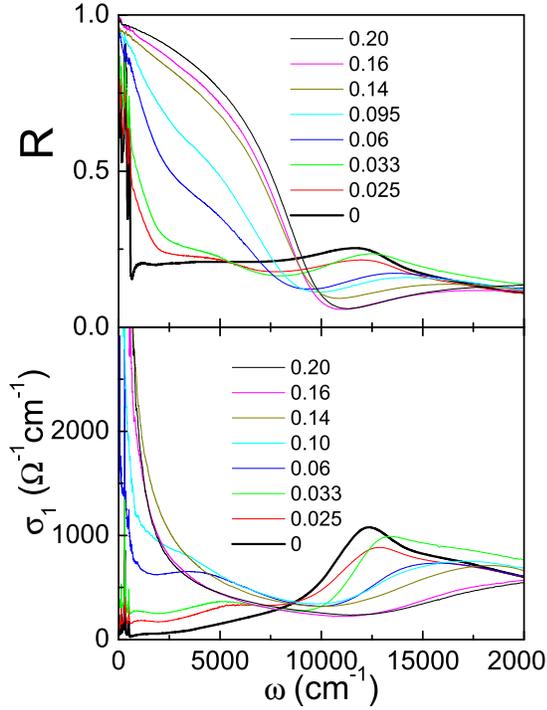}}%
\vspace{10pt}
\caption{(Color online) The evolution of in-plane
reflectance and
conductivity spectra for Nd$_{2-x}$Ce$_x$CuO$_4$ with doping.}%
\label{fig1}
\end{figure}

\section{Chemical potential shift with doping}

Figures. 1 shows the room temperature $R(\omega$) and conductivity
($\sigma_1(\omega$)) spectra for NCCO. The undoped $x$=0 crystal
shows a broad peak at 1.5 eV (12000 \cm) due to the
charge-transfer excitations. At low frequency, it has very low
conductivity values except for some infrared-active phonon lines.
Upon Ce substitution, a transfer of spectral weight from
charge-transfer excitations to low frequency occurs. A
mid-infrared broad peak at about 4000-5000 \cm is formed first at
low doping, then a Drude component appears at lower frequencies
with further doping. Such spectral change has been observed
previously \cite{Cooper,Arima,Onose} and shares many similarities
with the hole-doped cuprates \cite{Uchida}. However, one important
feature which was not addressed previously is that, accompanying
the spectral weight transfer, the charge-transfer excitation peak
also shifts gradually to higher frequencies with increasing
doping. This behavior is associated with the change of chemical
potential with electron doping, as elaborated below.

\begin{figure}
\centerline{\includegraphics[width=2.8in]{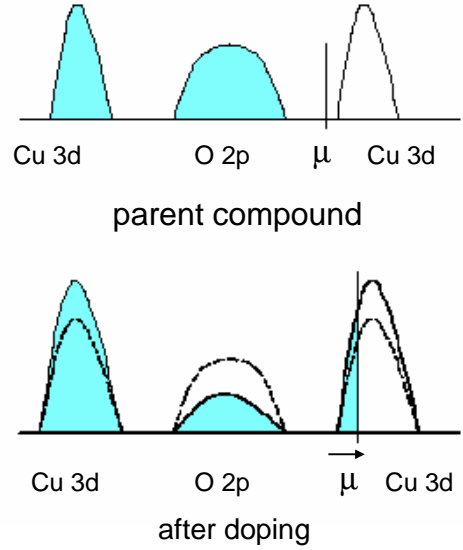}}%
\caption{(Color online) A schematic picture for the doping
evolution of electronic states of Nd$_{2-x}$Ce$_x$CuO$_4$. For
undoped parent compound, the chemical potential locates close to
the Cu 3$d$ upper Hubbard band. With electron doping, the chemical
potential moves into the conduction band and shifts up.}%
\label{fig1}
\end{figure}

It is useful to compare the above result with ARPES data. ARPES
measurement on undoped Nd$_2$CuO$_4$ revealed a dispersion band
$\sim$ 1.3 eV below the chemical potential $\mu$ along zone
diagonal, which was ascribed to the oxygen-derived charge transfer
band \cite{Armitage}. Note that this energy (1.3 eV) is lower than
the charge-transfer excitation ($\sim$1.5 eV) seen in optical
measurement. As we mentioned, the ARPES experiment detects the
occupied states relative to the chemical potential, while optics
probes the interband transition from an occupied band below $\mu$
to an unoccupied band above $\mu$. The small difference between
the two probes indicates that the $\mu$ locates just slightly
below the Cu 3$d$ upper Hubbard band in Nd$_2$CuO$_4$ (Fig.2). By
contrast, the $\mu$ for the hope-type parent compound
La$_2$CuO$_4$ is close to the oxygen 2$p$ band. This is because
the dispersive band seen in ARPES measurement is not far from the
Fermi level \cite{Yoshida}, while the charge-transfer excitations
probed by optical measurement are around 2 eV \cite{Uchida}. The
fact that $\mu$ is very close to the Cu 3$d$ upper Hubbard band in
Nd$_2$CuO$_4$ and to the O 2$p$ band in La$_2$CuO$_4$ explains why
electrons and hole are easily doped into those different types of
compounds, respectively.

There exists a long standing controversy over the evolution of
$\mu$ with doping in cuprates \cite{Damascelli}. One simple
picture is that $\mu$ moves into the valence or conduction band as
the material is doped away from half filling with holes or
electrons, respectively. Another opinion is that carrier doping
creates "states" inside the insulator's gap, but $\mu$ remains
relatively fixed in the middle of the gap. It is difficult to get
to a definite solution to this issue solely from APRES
measurement. In fact, the experimental data on different materials
have been interpreted in terms of both scenarios
\cite{Allen,Ino,Veenedal}. The present optical experiments,
showing a shift of charge-transfer excitation with doping, is
obviously consistent with the scenario that the $\mu$ enters the
upper Cu 3$d$ Hubbard band and shifts up with electron doping. As
illustrated in Fig. 2, when the chemical potential moves into the
valence band, some states in Cu 3d upper Habbard band were
occupied, then the charge-transfer transition from the oxygen 2$p$
to upper Cu 3$d$ Hubbard band would require a bit higher energy
than the charge-transfer gap. The chemical potential shift in NCCO
was also seen in a core-level photoemission study \cite{Harima}.

\section{Evolution of "hot spots" gap}

\subsection{Gap features in conductivity and scattering rate spectra}

Figure 3 shows the $T$-dependent $R(\omega)$ spectra for several
doping levels $x$=0.06, 0.10, 0.14, and 0.20, respectively. The
$x$=0.06 and $x$=0.10 samples are heavily underdoped and locate
below the superconducting region, the $x$=0.14 sample is slightly
underdoped with $T_c\approx$18 K, while the $x$=0.20 sample is
highly overdoped and also out of the superconducting region in the
phase diagram. The most prominent feature is a reverse S-shape at
low $T$ for the underdoped samples. Correspondingly, the
conductivity spectra, displayed in the upper panels of Fig. 4,
show suppressions of spectral weight below $\sim$0.5 eV (4000
\cm). Such behavior was observed in previous optical measurement
and referred to as a pseudogap \cite{Onose}. The suppressions are
clearly seen for $x$=0.06 and 0.10 samples, and are also present
in the low-$T$ curves for the $x$=0.14 superconductor. In the mean
time, Drude-like peaks are still seen at very low frequency for
all those samples, evidencing metallic responses in both $T$- and
$\omega$-dependences. Those observations indicate clearly that the
gap appears only on parts of the Fermi surface. The Drude-type
contribution originates from the gapless regions of the Fermi
surface.

\begin{figure}
\centerline{\includegraphics[width=3.2in]{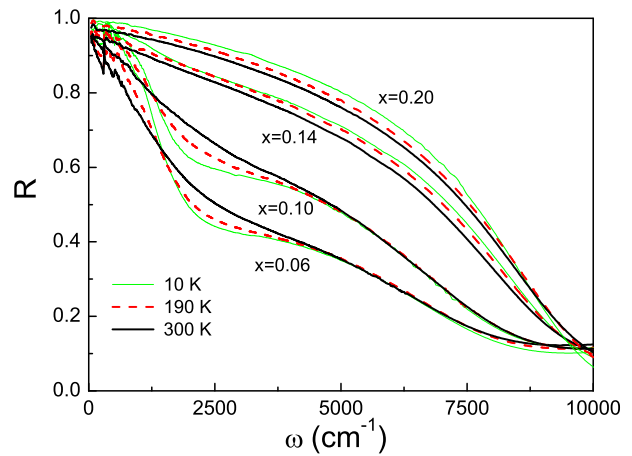}}%
\caption{(Color online) The temperature dependence of the
reflectance spectra for
several doping levels.}%
\label{fig2}
\end{figure}

The dynamics of charge carriers is usually described in terms of
frequency-dependent scattering rate on the basis of the extended
Drude model,
\begin{equation}
   {1\over\tau(\omega)}={\omega_p^2\over4\pi}Re{1\over\sigma(\omega)},
\label{chik}
\end{equation}
where $\sigma(\omega)$=$\sigma_1(\omega)$+$i\sigma_2(\omega)$ is
the complex conductivity, $\omega_p$ is the plasma frequency which
can be obtained by summarizing $\sigma_1(\omega)$ up to the
reflectance edge frequency. The scattering rate spectra for the
above four samples are shown in the bottom panels of Fig.~4.
Related to the suppression in $\sigma_1(\omega)$ spectrum, we
observe a peak in 1/$\tau(\omega)$ for underdoped
Nd$_{2-x}$Ce$_x$CuO$_4$ samples, below which the scattering rate
is strongly suppressed.

It should be noted that a depression of 1/$\tau(\omega)$ at low
frequencies followed by a peak or overshoot at high frequencies is
also a characteristic feature for a gaped state. To explain such
structure, Basov et al.\cite{Basov1} introduced an approximate sum
rule for the difference between energy-dependent scattering rates:
\begin{equation}
   \int_0^{\omega_c}d\omega({1\over\tau^A(\omega)}-{1\over\tau^B(\omega)})\simeq0.
\label{chik}
\end{equation}
where indexes A and B refer to different states of the studied
system (e.g., normal, pseudogap, superconducting). This sum rule
is based on the exact value of the imaginary part of the loss
function,
\begin{equation}
   \int_0^\infty{d\omega\over\omega}Im[\epsilon^{-1}(\omega,T)]=-{\pi\over2}.
\label{chik}
\end{equation}
In the case of $\omega<\omega_p$, the scattering rate can be
expressed as 1/$\tau(\omega)$$\approx\omega_p^{-1}Im(1/\epsilon)$
and therefore the sum rule is obeyed. According to equation (2),
the suppression of 1/$\tau(\omega)$ in the intragap region ought
to be balanced out by the overshoot at $\omega>E_g$.\cite{Basov1}
Indeed, such characteristic structure, \textit{i.e.} a decrease
and then an overshoot in 1/$\tau(\omega)$, was found to be present
in materials with development of different sorts of gaps. For
example, the behaviors were seen in antiferromagnet Cr where a
spin-density-wave (SDW) gap opens on parts of the Fermi surface
\cite{Basov1}, and in the heavy-fermion materials
YbFe$_4$Sb$_{12}$ and UPd$_2$Al$_3$ where hybridization gaps due
to the mixing of the $f$ band and the conduction band open on
parts of the Fermi surface \cite{Basov1,Dressel}.

\begin{figure*}
\centerline{\includegraphics[width=6.0in]{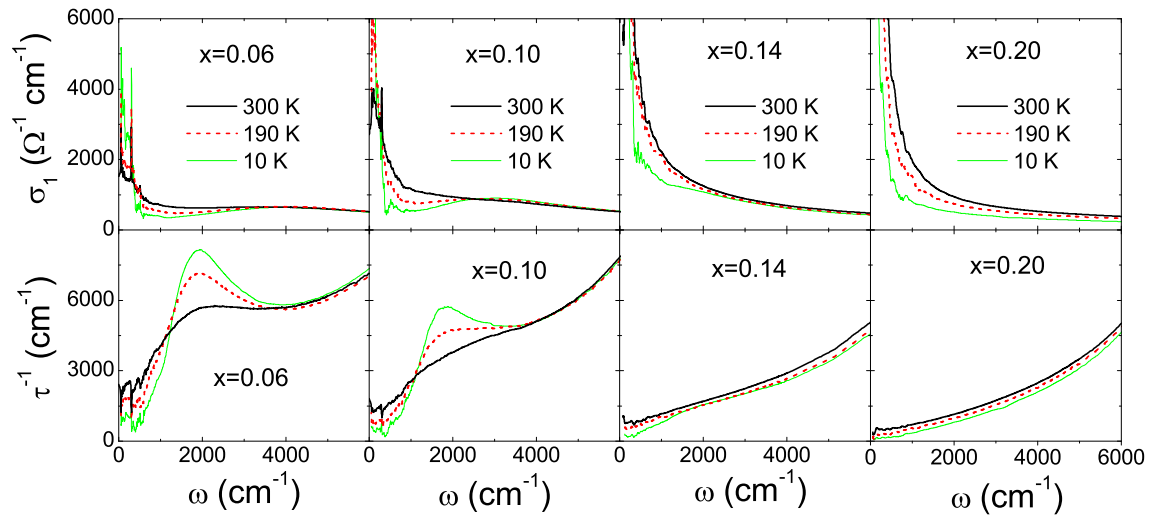}}%
\caption{(Color online) The temperature dependence of the
conductivity and
scattering rate spectra for $x$=0.06, 0.10, 0.14 and 0.20 samples.}%
\label{fig3}
\end{figure*}

In high-T$_c$ cuprates, the effect of the formation of a
superconducting gap or the development of a pseudogap on the
scattering rate is more complicated. Due to the fact that the
in-plane carrier dynamics is governed by the nodal region of the
Fermi surface that remains gapless at all temperature, the gap
effect is usually very weak. For optimally doped YBCO\cite{Basov1}
and Tl-based systems\cite{Wang}, a small overshoot shows up
immediately above the suppression in 1/$\tau(\omega)$ in the
superconducting state. However, for underdoped cuprates,
particularly in the pseudogap state, the overshoot is usually not
observed, the spectral weight lost at low $\omega$ is not
recovered up to the highest frequency measured. A through
discussion about the respective effects of superconducting gaps
with s-wave and d-wave symmetries on the sum rules for
1/$\tau(\omega)$ was provided by Marsiglio et al.\cite{Marsiglio}.
Nevertheless, for the electron-doped compounds under present
investigation, the 1/$\tau(\omega)$ spectra display significant
peak structures just above the suppressions. The energy scale of
the peak is much larger than that of the small overshoot below
T$_c$ in 1/$\tau(\omega)$ for hole-type cuprates at optimal
doping. Therefore, the structure is specific to electron-doped
cuprates. It provides strong evidence for the gap formation on the
Fermi surface in the underdoped NCCO. Apparently, the gaps in the
present case are not superconducting ones, but correspond to gaps
at "hot spots" with a much larger energy scale. Because the hot
spots locate more closely to the nodal region, the gap features
are seen more pronounced.

\subsection{Different energy scales for the peaks in $\sigma_1(\omega)$
and 1/$\tau(\omega)$ and their evolutions with doping}

We notice that the peak positions in 1/$\tau(\omega)$ spectra can
be quite different from the peak positions in $\sigma_1(\omega)$
spectra. For example, in the $x$=0.06 sample, the peak in
$\sigma_1(\omega)$ locates near 4000 \cm, while the peak in
1/$\tau(\omega)$ is only around 2000 \cm. A natural question is
which one reflects the gap amplitude? Previously, the pseudogap
observed in optics and its connection with the gap seen in ARPES
was established from $\sigma_1(\omega)$ spectra \cite{Onose}, but
this should be reexamined.

It is worth mentioning that there is no simple and direct
connection between the quantity $\sigma_1(\omega)$ or
1/$\tau(\omega)$ and the density of states (DOS). In order to know
the gap magnitude, one needs a model to derive those optical
quantities which, in general, involve a summation over the full
Fermi surface. As there is no theoretical study available for
distinguishing the gap features in the two optical quantities, we
consider a special case here. We calculate the conductivity
spectra for a standard BCS model with an energy gap of 2$\Delta$
in both clean and dirty limits for $T \ll T_c$,\cite{Zimmerman}
then determine 1/$\tau(\omega)$ from the extended Drude model
(equ. (1)). The model is applicable to both the superconducting
and the CDW/SDW gaps. The results are shown in Fig.~5. The
calculation well reproduces the sharp peak structure near
2$\Delta$ in 1/$\tau(\omega)$ as obtained by Basov \textit{et al}.
\cite{Basov1}, which is due to the presence of a singularity of
DOS. It indicates clearly that, for the same strength of impurity
scattering, the peak position in $\sigma_1(\omega)$ could be very
different from that in 1/$\tau(\omega)$, although both can occur
above the gap energy. The peak in $\sigma_1(\omega)$ always
appears at a higher energy. But the energy difference between the
peaks in $\sigma_1(\omega)$ and 1/$\tau(\omega)$ becomes smaller
with reducing the impurity scattering due to a faster shift of the
peak position towards the low frequency in $\sigma_1(\omega)$. On
this account, the gap energy is much closer to the peak energy in
1/$\tau(\omega)$ than that in $\sigma_1(\omega)$.

Although the calculations are for the case of a fully gapped Fermi
surface, the experimental data for the "hot spot" gaps of NCCO
(i.e. partial gaps in the Fermi surface) qualitatively follow the
trend both in the peak positions and with doping evolution. Since
the electrons in heavily underdoped samples experience stronger
scattering, a larger energy difference between those peaks was
observed, as one would expect. Additionally, the gap magnitude in
1/$\tau(\omega)$ is indeed comparable to the gap energy seen in
ARPES measurement. ARPES experiments have well established that
the partial energy gap locates at the intersecting points (hot
spots) of the Fermi surface with the AF zone boundary, pointing
towards a SDW-type origin due to the strong AF scattering. Matsui
\textit{et al}. recently measured the $T$-dependent ARPES spectra
on a NCCO crystal at $x$=0.13,\cite{Matsui} and identified a
maximum gap (i.e. a hump feature in an energy dispersion curve)
$\sim$0.19 eV. This value is very close to the peak positions in
1/$\tau(\omega)$ for the compounds with similar Ce contents.

\begin{figure}
\centerline{\includegraphics[width=2.7in]{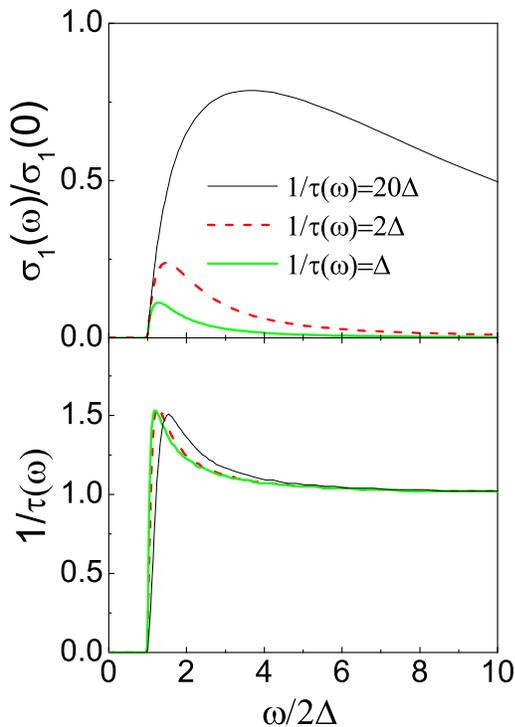}}%
\caption{(Color online) The calculated conductivity and scattering
rate spectra for a BCS model ($T \ll T_c$) in both the clean and
dirty limits with (1/$\tau)/\Delta$=1, 2, and 20.}%
\label{fig3}
\end{figure}

We now address the evolution of the AF gap with doping. From Fig.
4, we found that the peak in 1/$\tau(\omega)$ shows much less
frequency shift than the suppression feature in $\sigma_1(\omega)$
with increasing doping. The most prominent change is a weakening
of the gap structure with increasing doping and temperature. The
results suggest that the disappearance of the gap feature is not
due to the reduction of the gap energy, but in a manner of
gradually fading away. The result is well consistent with Raman
scattering data on the two-magnon peak: its intensity decreases
with Ce doping but the peak energy shows little doping dependence
\cite{Onose}. ARPES data on the underdoped
Nd$_{1.87}$Ce$_{0.13}$CuO$_4$ further indicated that the AF
pseudogap was gradually filling up with increasing T without
showing any observable shift in energy \cite{Matsui}. Note that
although the gap is due to the AF scattering \cite{Tohyama}, it
opens at a temperature much higher than the AF ordered Neel
temperature $T_N$ or in a doping region where there is no long
range AF ordering. Clearly, the gap already shows up when the
short range AF correlation exists. Unlike the AF long range order
which has a definite phase transition temperature, the AF short
range order is a gradually evolving behavior with no phase line as
a function of doping. Some authors suggest the existence of a
quantum phase transition in the phase diagram for electron-doped
cuprates \cite{Zimmers,Dagan}. However, our analysis above
suggested that the gap feature is associated with the short range
AF interaction, with no clear boundary line or any critical point.

Another important feature is that, when the pseudogap goes away by
further increasing doping, the 1/$\tau(\omega)$ indicates a
dominant $\omega^2$-dependence (see the 1/$\tau(\omega)$ spectra
for the $x$=0.20 samples in Fig.~4), suggesting a Fermi
liquid-like state. The 1/$\tau(\omega)$ spectra of the $x$=0.14
sample deviate from such $\omega^2$-dependence at the low
frequency even at room temperature, suggesting that the
short-range AF correlation is still effective at this doping level
and temperature range. We believe that, at heavily overdoped
region, due to the disappearance of the AF correlation, the
magnetic folding of the Fermi surface vanishes completely,
consequently only a single large Fermi surface centered at
($\pi,\pi$) is formed. The amazing observation is that this
$\omega^2$-dependence goes up to the very high frequency, beyond
6000 \cm as shown in Fig.~4. This is very different from the hole
doped cuprates. In the hole doped case, the scattering rate
follows a linear $\omega$-dependence which is well described by
the Marginal Fermi liquid theory, whereas at high frequencies
(around 0.5 eV) a saturation of 1/$\tau(\omega)$ is commonly
observed \cite{Marel}.

\section{Summary}

Our systematic study on the NCCO crystals reveals a number of
novel properties: (1) The chemical potential is not pinned in the
gap region, but shifts up with electron doping. (2) Both
$\omega$-dependent conductivity and scattering rate spectra
exhibit characteristic structures caused by "hot spots" gaps in
the Fermi surface for underdoped samples. The peak position in
1/$\tau(\omega)$ is found to be quite different from the peak
position in 1/$\sigma(\omega)$. We clarify that the "hot spots"
gap probed by ARPES experiments correlates directly with the
suppression feature in 1/$\tau(\omega)$ rather than in optical
conductivity spectra. (3) The gap becomes weaker with increasing
doping/temperature and disappears in a manner of fading away.
There is no clear phase transition line for the gap vanishing. (4)
In the heavily overdoped region, the gap structure disappears
completely due to the absence of the AF correlation, and a single
large Fermi surface should be formed. We identify a
$\omega^2$-dependence of 1/$\tau(\omega)$ up to very high energy
with no indication of saturation.

\section{Acknowledgement}

We acknowledge helpful discussions with D. N. Basov, J. L. Luo, T.
Timusk, T. Tohyama, Z. Y. Weng, and T. Xiang. This work was
supported by National Science Foundation of China, the Knowledge
Innovation Project of Chinese Academy of Sciences, and  the
Ministry of Science and Technology of China£¨973 project
No:2006CB601002). HD acknowledges the support from US NSF
DMR-0353108.

%

\end{document}